# Monolithically integrated waveguide-coupled silica microtoroids

J. Richter[*,1], M. P. Nezhad[1,2], and J. Witzens[1]

[1]Institute for Integrated Photonics, RWTH Aachen, 52074 Aachen, Germany
[2]Now at Bangor University, School of Electronic Engineering, Dean Street, Bangor LL57 1UT, UK
[*]jrichter@iph.rwth-aachen.de

*Abstract*—We report on the design and fabrication of a new type of microtoroid high-Q silica resonators monolithically coupled to on-chip silicon nanowire waveguides. In order to enable monolithic waveguide coupling, the microtoroid geometry is inverted such that the resonator is formed by thermal reflow at the circumference of a hole etched in a suspended $SiO_2$ membrane. This configuration is shown to be conducive to integration with a fully functional Silicon Photonics technology platform.

## I. Introduction

Microresonators shaped by surface tension minimization of a melted silica film (microtoroids) are known for their extraordinarily high quality factors (Q-factors) on the order of a 100 Million [1], which enable the generation of frequency combs via nonlinear effects with very low thresholds [2]. The originally proposed microtoroids are typically coupled to an external tapered fiber [1]. Existing integration schemes of microtoroid type structures with on-chip waveguides have either relied on silica waveguides [3] or on suspended silicon (Si) waveguides fabricated in stacked silicon-on-insulator (SOI) device layers with a rather complex fabrication process [4]. Here we introduce a device geometry allowing straightforward coupling with Si nanowire waveguides and integration with existing Silicon Photonics (SiP) technology, monolithically combining microtoroids with devices such as electro-optic modulators, wavelength multiplexers, grating couplers or germanium photodetectors [5]. The main concept consists in combining a Si waveguide fabricated in the device layer of a SOI chip with an inverted version of a microtoroid fabricated by reflowing the buried oxide (BOx) of the same chip. Even though the inverted geometry results in a weakened confinement, simulations indicate that such a structure is able of achieving comparably high Q-factors as conventional microtoroids. A balance has however to be struck in choosing the thickness of the BOx layer by trading off Si waveguide substrate coupling losses against microtoroid bending losses.

## II. Inverted Microtoroid Fabrication

The fabrication starts with the definition of fully etched Si waveguides in the device layer of a SOI wafer by electron beam lithography (EBL) followed by cryogenic (-130°C) reactive ion etching (RIE) with $SF_6$. A thin $SiO_2$ layer (40 nm) is then formed on the waveguides' surface by rapid thermal processing (RTP) to protect them from a subsequent $XeF_2$ etch.

Next, a circular opening is defined in the BOx layer with optical lithography followed by a $CHF_3$ oxide etch reaching down to the Si substrate. A gas phase $XeF_2$ etch is then used to isotropically undercut the BOx layer by a distance $d_1$. In a post-processing step, the actual inverted microtoroid is created by melting the resulting freestanding circular $SiO_2$ membrane with a continuous wave (CW) $CO_2$ laser (20 W at λ=10.6 µm). Selective absorption of the laser light by the membrane results in melting and reflow of its circumference into a toroid shaped structure with a minor (inner) radius $r_{minor}$ and a major (outer) radius $r_{major}$ (Fig. 1). As with prior microtoroid type resonators, surface tension minimization during the reflow process results in minimal surface roughness, allowing for the ultra-high targeted Q-factors. Since the reflow of the $SiO_2$ membrane is a self-limiting process due to enhanced heat-sinking as the toroid approaches the edge of the undercut region, with the reflow process significantly slowing down near the edge of the undercut, the distance $d_2$ between the toroid and the latter can be relatively well controlled. This also allows setting the gap between the microtoroid and the integrated Si waveguide. In our current setup we are able to control it with a precision better than ±1µm limited by the imaging resolution, as well as varying device designs and overall process control (in particular the exact depth of the undercut).

In the final device, the Si waveguide in the coupling section lies on the remaining unmelted freestanding $SiO_2$ film and can

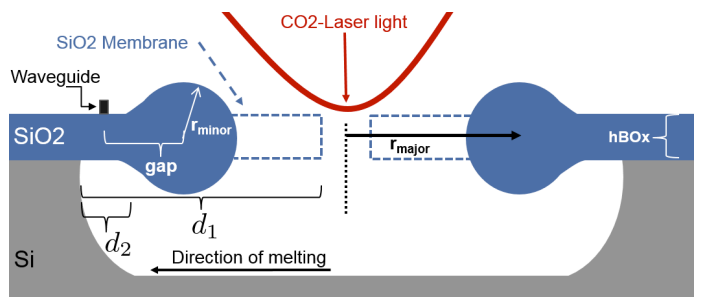

Fig. 1. Schematic cross-section of the inverted microtoroid illustrating the reflow process. The $SiO_2$ membrane forms an inverted toroid when melted. Since the distance between the center of the microtoroid and the rim of the undercut decreases as the reflow progresses, heat-sinking via the Si substrate is enhanced and the melting process slows down.

The authors gratefully acknowledge support from the European Research Council under contract 279770 and by the European Union's Seventh Framework Program under contract 293767.

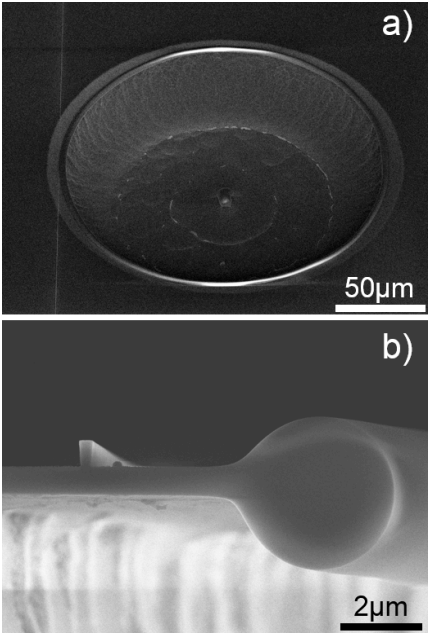

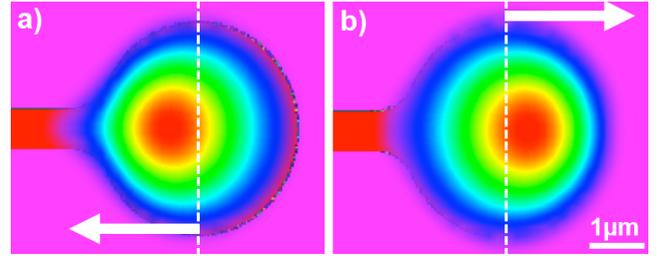

Fig. 3. Comparison of the $TM_0$ mode profiles for an inverted (a) and a conventional microtoroid (b) for $h_{BOx}$=0.8 µm, $r_{minor}$=2.1 µm and $r_{major}$= 125 µm. The white arrows indicate the direction of light leakage.

Fig. 2. Scanning electron microscope (SEM) images of a waveguide coupled inverted microtoroid. (a) Overview of the device (waveguide to the left) and (b) cross-section of a cleaved device together with the coupling silicon waveguide.

be routed away to the rest of the chip. A micrograph of a fabricated device is shown in Fig. 2. It was fabricated out of an earlier SOI material with a 700 nm BOx layer and a 400 nm device layer. In order to resolve quality issues resulting in waveguide delamination, we have since moved to a higher quality industry grade material with an 800 nm thick BOx and a 300 nm device layer. The reported simulations also focus on this latter material system.

Particular care has to be taken while optimizing the $XeF_2$ etch. A smooth etch is crucial as excessive roughness at the rim of the undercut is transferred to the shape of the microtoroid during reflow due to the heat-sinking mediated process self-limiting. This can spoil the Q-factor.

### III. MODELING

In order to identify suitable device geometries, simulations were performed using Synopsys' design suite RSOFT. The effects of BOx layer thickness $h_{BOx}$ on bending losses and on waveguide substrate coupling losses were first investigated with mode solves in cylindrical coordinates relying on the finite-elements method (FEM). After selecting a waveguide and microtoroid geometry, the Si waveguide to microtoroid coupling strength was investigated with three dimensional finite-difference time-domain (FDTD) simulations.

#### A. Bending Losses

The primary difference between a conventional and an inverted microtoroid resides in a much weaker mode confinement towards the outer region of the microtoroid's circumference for the latter. In the case of a conventional microtoroid the mode is pressed towards the outer rim of the core located opposite to the suspended $SiO_2$ film, with a high dielectric contrast between $SiO_2$ and air. On the other hand, in the case of an inverted microtoroid, the mode is pressed in the direction of the membrane, leading to increased bending losses due to leakage into membrane modes, which have a higher effective index than the refractive index of air (Fig. 3).

Based on volume conservation, the microtoroid core radius $r_{minor}$ is a function of the BOx thickness $h_{BOx}$ and the width of the reflown film. The core radius can be simply expressed as

$$(d_1-d_2)h_{BOx}/\pi = r_{minor}^2 \quad (1)$$

Fig. 4 shows the maximum loaded Q-factor $Q_{Bend}$ as a function of $h_{BOx}$ assuming critical coupling and a major radius of 125 µm (solid line) and 150 µm (dashed line), as limited by bending losses. Since we want the Q-factor to be primarilly limited by material properties such as absorption and scattering due to residual impurities and surface roughness, a $Q_{Bend}$ in excess of a 100 Million is targeted. This implies the BOx thickness has to be smaller than 0.9 µm (1µm for $r_{major}$ = 150 µm) for the transverse electric ($TE_0$) and smaller than 1.1 µm (1.3 µm for $r_{major}$ = 150 µm) for the transverse magnetic ($TM_0$) ground modes.

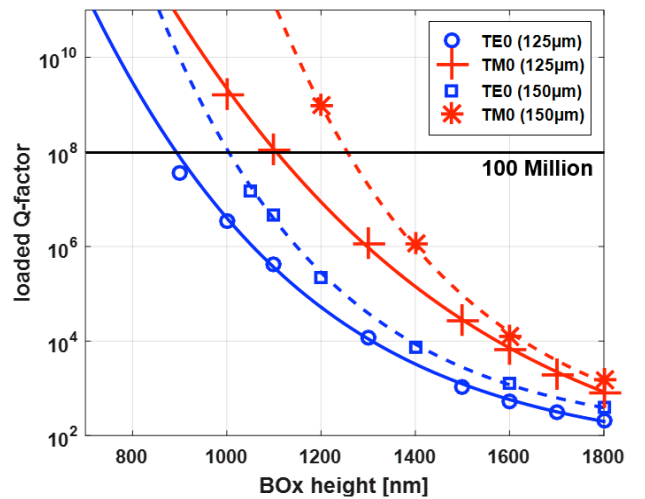

Fig. 4. Bending loss limited loaded Q-factor as simulated for $TE_0$ and $TM_0$ modes as a function of the BOx thickness. The core radius $r_{minor}$ is implicitly varied according to eq. 1. Solid lines correspond to a major radius of 125 µm and dashed lines to a major radius of 150 µm. The width and the height of the reflow region are assumed to be $d_1-d_2$=17 µm and $h_{BOx}$=800 nm.

## B. Waveguide Losses

Since the goal pursued here is the integration of microtoroids with a fully functional SiP technology platform, waveguide substrate coupling losses should be compatible with waveguide routing over several centimeters. Due to the comparatively thin BOx layer, the waveguide modes can couple to the Si substrate. For a waveguide width of 500 nm and above these coupling losses become small even for the $TM_0$ mode (Fig. 5). The toroid – waveguide coupling section is located on an undercut section of the $SiO_2$ film. In this case there are no substrate coupling losses. However, since the waveguide width in the coupling section is chosen to verify phase matching with the microtoroid, and for the waveguide to thus yield an effective index close to the refractive index of $SiO_2$, coupling of the waveguide mode into the BOx can occur. As a matter of fact, we taper the waveguide down to approximately 150 nm for coupling between $TM_0$ modes, at which point the effective index of the Si waveguide is already slightly below the index of $SiO_2$. Because the coupling section is quite short (<50 μm) the additional loss is small (<0.01 dB).

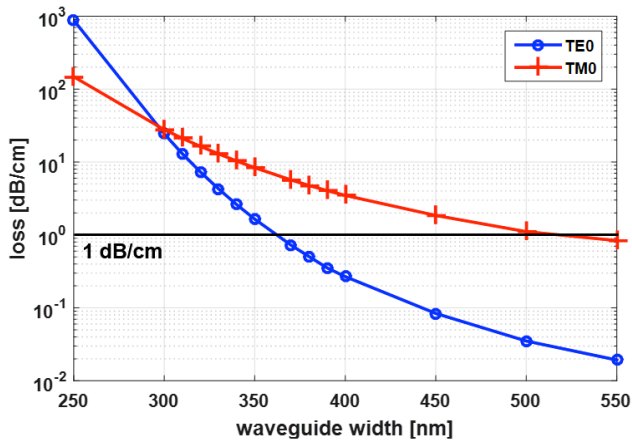

Fig. 5. Substrate coupling losses of a waveguide with a height of 280 nm and a 40 nm $SiO_2$ overcladding (resulting from the RTP oxidation) as a function of the Si waveguide core width for an 800 nm thick BOx layer on Si.

## C. Coupling Efficiency

In order to maximize optical power build-up inside the toroid and minimize the threshold for parametric wavelength conversion, it is important to reach a coupling point at or close to critical coupling. For conventional microtoroids this can be achieved by adjusting the gap between a tapered fiber and the microtoroid. In an integrated solution the gap is fixed and should in principle be well targeted by design and fabricated on the spot. In practice, however, the coupling strength can be corrected to some extent by tuning the wavelength or temperature. Since the waveguide and microtoroid are made out of different materials, one of which, Si, has a substantial thermo-optic coefficient, thermal tuning of the coupling strength is facilitated. Moreover, due to the strong dispersion of the tapered Si waveguide and wavelength dependent mode size, changing the wavelength can also serve to tune the coupling strength (Fig. 6). As an additional refinement, the waveguide – microtoroid junction can be replaced by two junctions in Mach-Zehnder interferometer configuration to facilitate tuning of the coupling strength at a fixed wavelength.

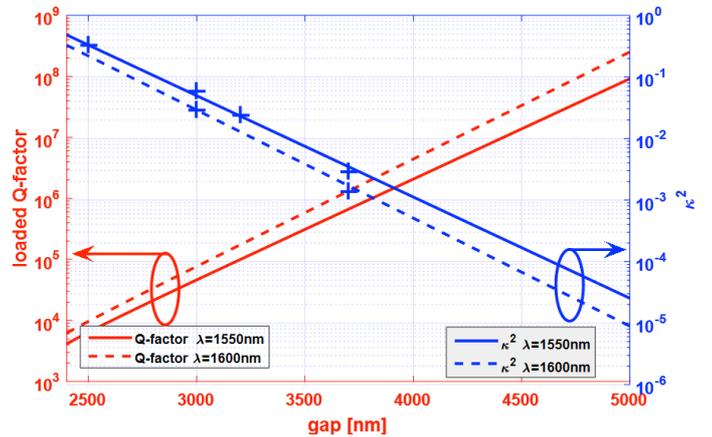

Fig. 6. Waveguide to microtoroid power coupling factor ($\kappa^2$) as a function of the gap (distance from waveguide to toroid core, center to center) for different wavelengths and for the $TM_0$ mode, as simulated with 3D FDTD. The corresponding loaded Q-factor assuming critical coupling (i.e., assuming the scattering and absorption limited intrinsic Q-factor to be twice the value) is also shown. The assumed structural parameters are $r_{minor}$=2.1 μm, $r_{major}$=125 μm and $h_{BOx}$=0.8 μm. The Si waveguide is assumed to be 150 nm wide in the coupling section, 280 nm high, and to be covered by 40 nm of $SiO_2$.

## IV. OUTLOOK ON BACK-END INTEGRATION

Integration of inverted microtoroids as described here with a completely functional SiP technology requires compatibility with a back-end stack. While the microtoroid can in principle be formed by reflowing a $SiO_2$ film formed by stacking the BOx layer with the back-end dielectrics, this results in a number of complications: For one, the effective index of the tapered waveguide will increase due to the upper oxide cladding, making phase matching more challenging. Moreover, the lower purity of the back-end dielectrics might lead to excess absorption losses inside the microtoroid. Finally, as shown in Fig. 4, the total thickness of the film is limited in order to maintain a high Q-factor. Maintaining a thin back-end of at most a few hundred nanometers can alleviate all three aspects. This could be achieved by locally thinning the back-end in the microtoroid region. Implementing this with a sacrificial layer that can be selectively chemically removed as the last step of the back-end thinning process (such as a silicon nitride layer) might be a good way to realize this with minimal additional fabrication variability and excess roughness.

## V. CONCLUSIONS

Simulations indicate that the monolithic integration of ultra-high-Q inverted microtoroids in a fully functional SiP technology platform is feasible. We have developed a suitable fabrication flow and fabricated first devices. While we have not yet realized the expected Q-factors, we expect this to be due to the insufficient material quality of the first set of custom made SOI material we have been working with. We are currently working on duplicating these results with high-grade SOI wafers.